\documentclass{elsarticle}

\usepackage[dvips]{graphicx}
\usepackage{amssymb}
\usepackage{subfig}
\usepackage{color}
\usepackage{alphalph}

\newcommand{\pt}{$p_T$}
\newcommand{\zt}{$z_T$}

\newcommand{\deta}{$\Delta\eta\;$}
\newcommand{\ptassoc}{$p_{T}^{assoc}\;$}
\begin{document}
\begin{frontmatter}



\title{System size dependence of associated yields in hadron-triggered jets}

\author[UoIC]{B.~I.~Abelev}
\author[Panjab]{M.~M.~Aggarwal}
\author[Kolkata]{Z.~Ahammed}
\author[Kent]{B.~D.~Anderson}
\author[ParticleDubna]{D.~Arkhipkin}
\author[HighDubna]{G.~S.~Averichev}
\author[MIT]{J.~Balewski}
\author[UoIC]{O.~Barannikova}
\author[Birm]{L.~S.~Barnby}
\author[Strasbourg]{J.~Baudot}
% ...
\address{\textsc{(STAR Collaboration)}}

% ...
\address[Birm]{University of Birmingham, Birmingham, United Kingdom}
% ...
\address[UoIC]{University of Illinois at Chicago, Chicago, Illinois 60607, USA}
% ...
\address[HighDubna]{Laboratory for High Energy (JINR), Dubna, Russia}
\address[ParticleDubna]{Particle Physics Laboratory (JINR), Dubna, Russia}
% ...
\address[Strasbourg]{Institut de Recherches Subatomiques, Strasbourg, France}
% ...
\address[Kent]{Kent State University, Kent, Ohio 44242, USA}
% ...
\address[MIT]{Massachusetts Institute of Technology, Cambridge, MA 02139-4307, USA}
% ...
\address[Panjab]{Panjab University, Chandigarh 160014, India}
% ...
\address[Kolkata]{Variable Energy Cyclotron Centre, Kolkata 700064, India}
% ...

\begin{abstract}
We present results on the system size dependence of high transverse
momentum di-hadron correlations at $\sqrt{s_{NN}}$~=~200~GeV as measured
by STAR at RHIC. Measurements in d+Au, Cu+Cu and Au+Au collisions
reveal similar jet-like correlation yields at small angular separation
($\Delta\phi\sim0$, $\Delta\eta\sim0$) for all systems and
centralities. Previous measurements have shown that the away-side yield is suppressed in heavy-ion collisions. We present measurements of the
away-side suppression as a function of transverse momentum and
centrality in Cu+Cu and Au+Au collisions. The suppression is found to
be similar in Cu+Cu and Au+Au collisions at a similar number of
participants. The results are compared to theoretical calculations based on the
parton quenching model and the modified fragmentation model. The observed differences between data and theory indicate that the correlated yields presented here will provide important constraints on medium density profile and energy loss model parameters. 
\end{abstract}

\begin{keyword}
parton energy loss \sep jet quenching \sep di-hadron fragmentation function \sep relativistic heavy-ion collisions

\PACS 25.75.-q \sep 25.75.Gz
\end{keyword}
\end{frontmatter}

One of the important early results from the experiments at
RHIC is the observation of jet quenching in heavy-ion
collisions. The suppression of high transverse momentum ($p_{T}$)
particle production in inclusive hadron spectra~\cite{Adler:2003kg,Adams:2003am} and jet-like structures in di-hadron correlation measurements~\cite{Adler:2002tq} indicates that
partons originating from hard scatterings in the initial stages of the collisions
interact strongly with the created medium.

Previous studies~\cite{Dan} investigated azimuthal correlations of high-$p_{T}$ hadrons and showed that the suppression of the
correlated away-side yield increases with centrality in Au+Au collisions. Various theoretical calculations~\cite{Loizides2,XNian2,GLV,WDGH} of partonic energy loss in the medium have been used to derive
medium properties like the transport coefficient $\hat{q}$. The energy loss of individual partons is also expected to depend on the path length through the medium in a way that is
characteristic of the energy loss mechanism. 
For radiative energy loss, which is thought to be dominant for
light quarks, the energy loss is expected to depend on $L^{2}$,
the square of the traversed path length, due to coherence effects~\cite{BDMPS1,BDMPS2,BDMPS3}. 
For elastic energy loss, on the other
hand, a linear dependence on $L$ is expected. Prior results from RHIC have not established in detail the energy loss mechanism. Combined measurements of single-hadron and di-hadron suppression are sensitive to the path length dependence and can help
determine which process dominates~\cite{Renkel}. 
In addition, different implementations of the energy loss calculation use different
path-length distributions and density profiles. The system-size
dependence of away-side suppression is sensitive to these modeling
parameters and will provide further constraints~\cite{RenkC75}.


We present a systematic study of the near- ($\Delta\phi\sim0$) and
away-side ($|\Delta\phi|\sim \pi$) di-hadron correlated yields as a
function of the number of participant nucleons ($N_{part}$). Data for
three systems with different geometries (d+Au, Cu+Cu and Au+Au) at
$\sqrt{s_{NN}} = 200$ GeV were collected by the STAR experiment at
RHIC. A study of the hadron-triggered fragmentation functions in the three
systems is also presented. Results from d+Au collisions are used as a
reference without a hot medium. The d+Au data sample is preferred over the
p+p data sample because it has significantly larger
statistics. Earlier comparisons between p+p and d+Au collisions have
established that jet suppression is a final state effect and is not
present in d+Au collisions~\cite{dAu1,dAu2,dAu3,dAu4}.

\begin{figure*}\label{nearcorr}
\subfloat[]{\label{DeltaPhiNew1}
  \includegraphics[height=45mm,width=0.32\linewidth]{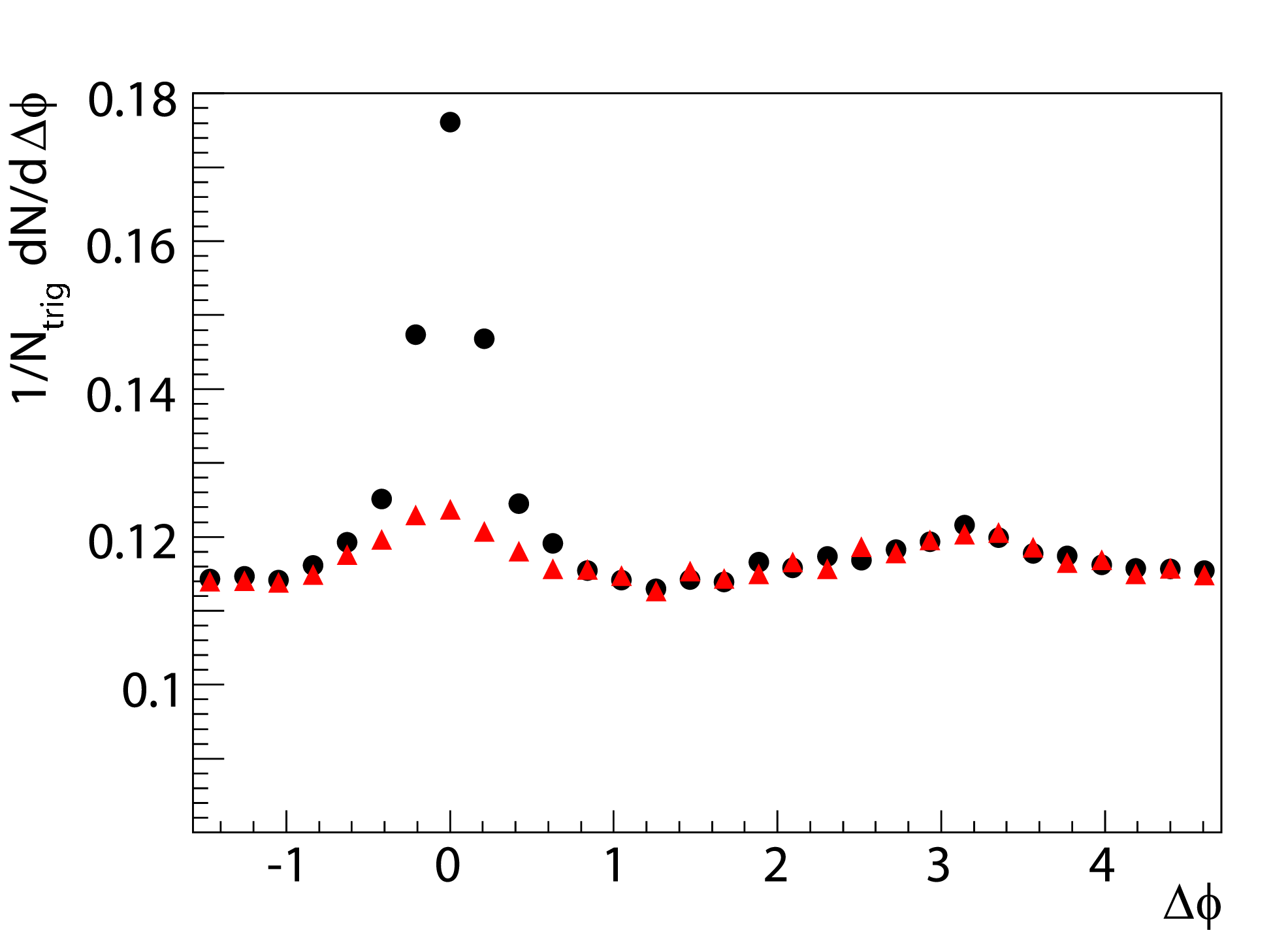}
  }
\subfloat[]{\label{DeltaPhiNew2}
  \includegraphics[height=45mm,width=0.32\linewidth]{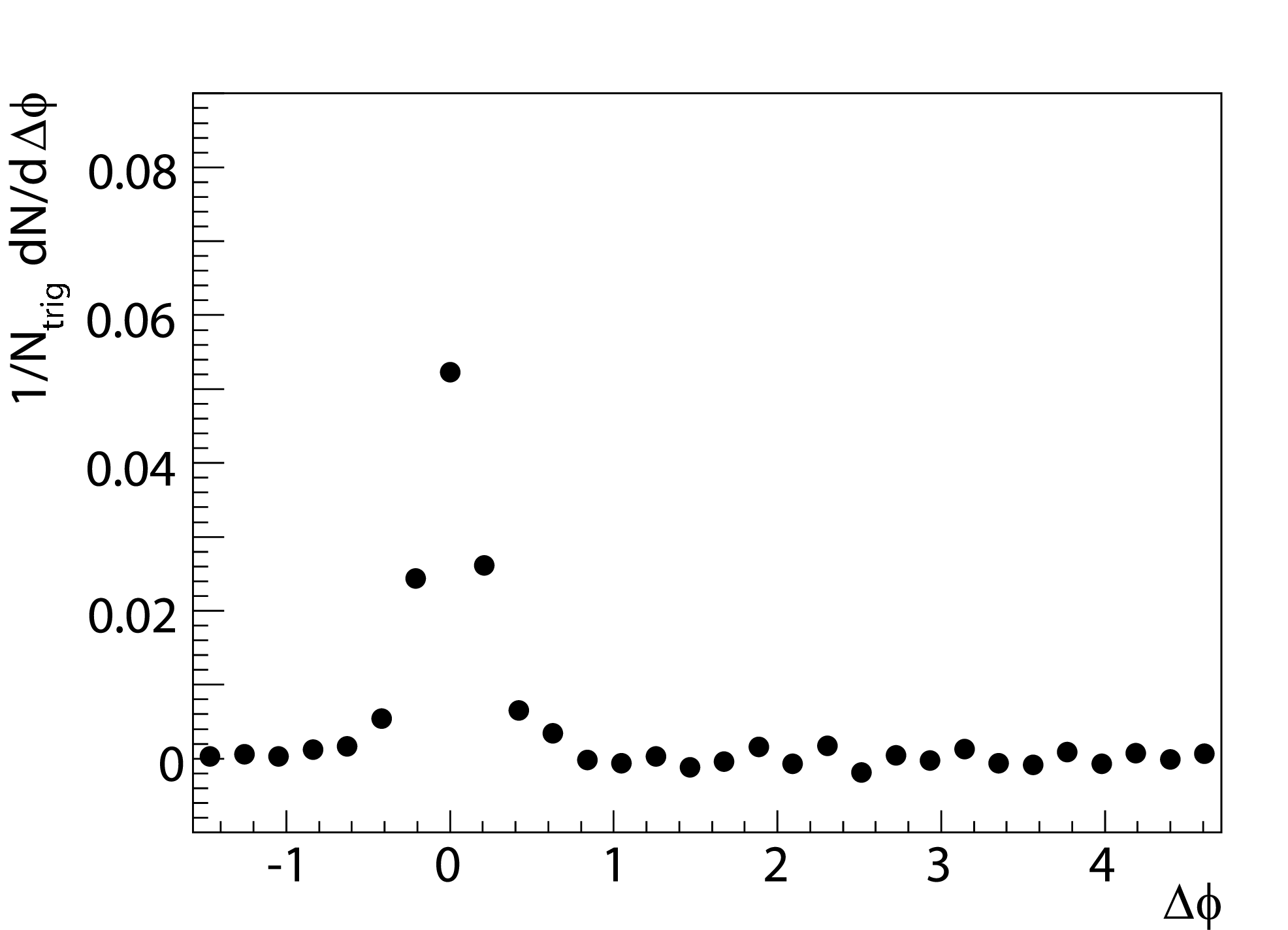}
  }
\subfloat[]{\label{DeltaPhiNew3}
  \includegraphics[height=45mm,width=0.32\linewidth]{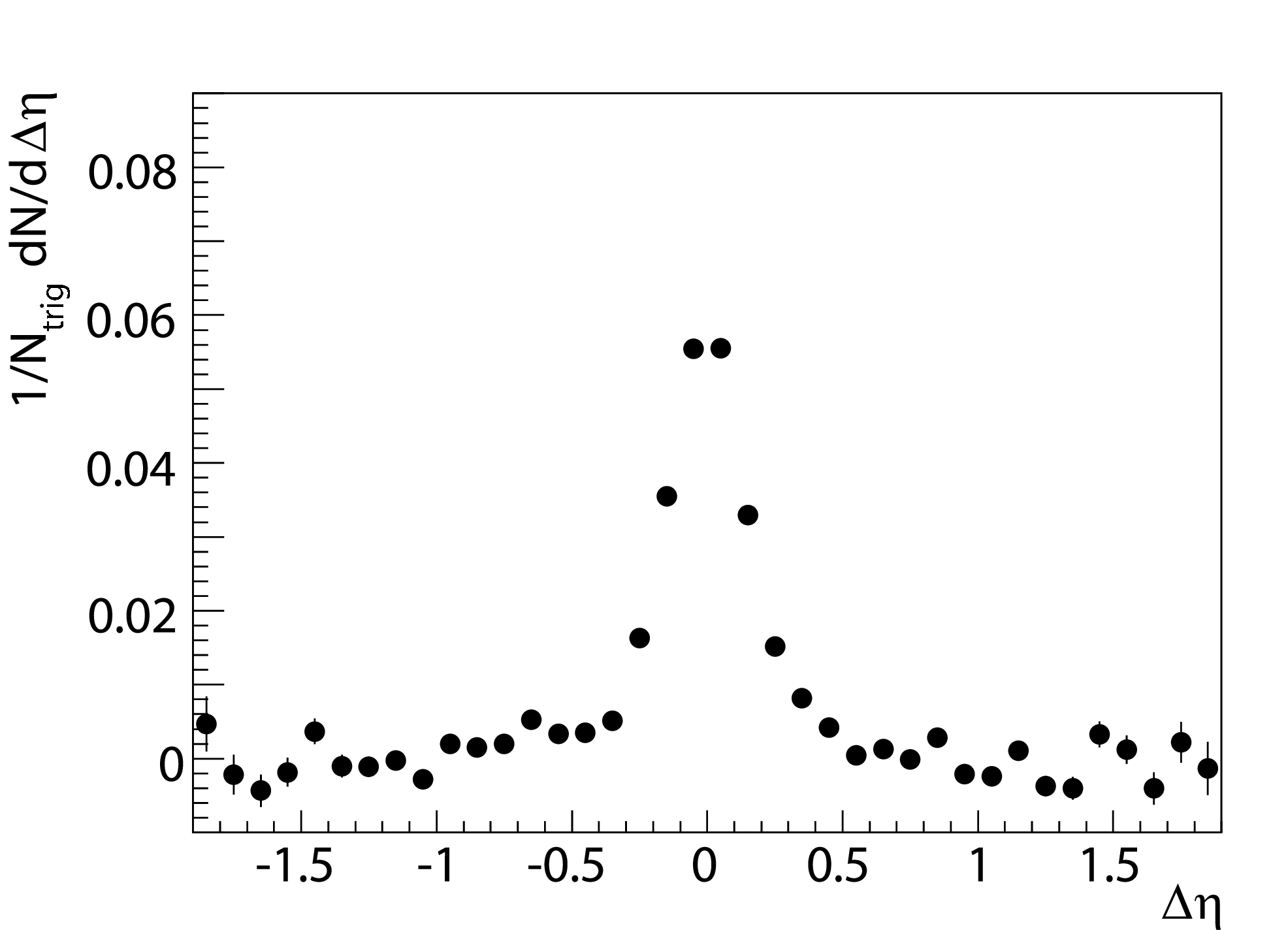}
} \caption{Di-hadron correlations in central (0-12\%) Au+Au collisions: (a) $\Delta\phi$ correlations - small $\Delta\eta$ ($|\Delta\eta|<0.7$) (black circles) and 
large $\Delta\eta$ ($0.7<|\Delta\eta|<1.7$) scaled to match the small $\Delta\eta$ result at large $\Delta\phi$ (red triangles), (b)
$\Delta\phi$ subtracted distribution,  (c) $\Delta\eta$ subtracted
distributions; 4~$<p_{T}^{trig}<$~6~GeV/$c$, $p_{T}^{assoc}>$3~GeV/$c$.}
\label{Cu_Cu_10}
\end{figure*}

This analysis is based on four data sets and includes 11.7
million minimum-bias d+Au events, 43.8 million minimum-bias Cu+Cu events, 25 million minimum-bias Au+Au events and 19 million Au+Au events collected using a central trigger. The central trigger uses the coincidence of two Zero Degree Calorimeters (ZDCs) and a multiplicity threshold in the Central Trigger Barrel~\cite{star} which selects the most central $0-12\%$ of total geometric cross-section. In order to minimize the influence of the Time Projection
Chamber (TPC) acceptance ~\cite{TPC_NIM}, only events with a reconstructed collision vertex with $|z_{vertex}|\leq30$~cm are
included in the analysis, where $z_{vertex}$ represents the distance along the beam-line from the center of the detector. 

The di-hadron correlations are formed using charged particles
reconstructed in the TPC, within a pseudorapidity range of
$-1<\eta<1$. High $p_{T}$ trigger particles are selected and the
$\Delta\eta\times\Delta\phi$ distribution of associated particles
(${p_{T}}^{assoc} < p_{T}^{trig}$) is constructed. An $\eta$, $p_{T}$
and centrality dependent reconstruction efficiency correction is
applied to obtain the associated particle yields. It is not necessary
to apply the efficiency correction to trigger particles when
calculating the correlated yields because the final result is normalised per trigger particle. The track reconstruction efficiency depends on the track
density within the TPC and ranges from 89\%
(peripheral collisions) to 77\% (central). The systematic uncertainty
on the efficiency correction is estimated to be 5\% and is strongly
correlated across centralities and $p_{T}$ bins for each data set, but
not between data sets.  
particles is not necessary since

Pair acceptance corrections in $\Delta\phi$ and $\Delta\eta$ are
computed using a mixed event technique. These corrections reflect the conditional
probability of reconstructing two tracks with a specified relative
kinematics. The dominant feature in the $\Delta\phi$ pair acceptance
correction are the small gaps between the sectors of the TPC. The
$\Delta\eta$ pair acceptance correction is of triangular shape, with
a maximum of 1 at mid-rapidity and minimum of 0 at the limit of the
pair acceptance $\Delta\eta=\pm2$. The $\Delta\eta$ acceptance correction is not
applied to the away-side yields.

Earlier results from STAR~\cite{ridge1,ridge2} have shown that
there is a finite associated yield on the near-side ($\Delta\phi\sim0$)
with large pseudorapidity separation $\Delta\eta$ (the ``ridge"). Since
the ridge properties are similar to those of the medium, 
it is appropriate to subtract this contribution in the present
analysis. In order to extract the jet contribution to the near-side
yield, the azimuthal correlation distribution for large $\Delta\eta$
separation ($0.7<|\Delta\eta|<1.7$) is subtracted from the
distribution for small $\Delta\eta$ ($|\Delta\eta|<0.7$). To account for the different $\Delta\eta$ window widths, the former distribution
is scaled so that the two distributions match in the away-side
region. This subtraction removes the \deta-independent ridge
contribution and the contributions from elliptic flow $v_2$, which is also
largely independent of $\eta$ in the range considered \cite{Adler:2002pu}. 
Figure~\ref{DeltaPhiNew1} shows central Au+Au distributions in the large (black) and small (red) $\Delta\eta$ regions, for trigger
particles with transverse momentum 4~GeV/$c$~$<{p_{T}}^{trig}<$~6~GeV/$c$,
and associated particles with ${p_{T}}^{assoc}$ in the range 3~GeV/$c$~$<{p_{T}}^{assoc}<{p_{T}}^{trig}$. The signal distribution after
subtraction is shown in Fig.~\ref{DeltaPhiNew2}. 

An alternative way to extract the near-side associated yield is to use
the $\Delta\eta$-distribution, which is obtained by projecting the
$\Delta\eta\times\Delta\phi$ correlations in the $|\Delta\phi| < 0.78$ region
onto the $\Delta\eta$ axis. In this projection, the
ridge yield and elliptic flow constitute a flat background which is determined by averaging the yield at large $|\Delta\eta|>0.7$ and 
subtracted. Figure~\ref{DeltaPhiNew3} shows a background subtracted
$\Delta\eta$ projection with the same trigger conditions as Fig.~\ref{DeltaPhiNew1} and Fig.~\ref{DeltaPhiNew2}.

\begin{figure*}
\subfloat[]{
\includegraphics[height=55mm,width=0.5\linewidth]{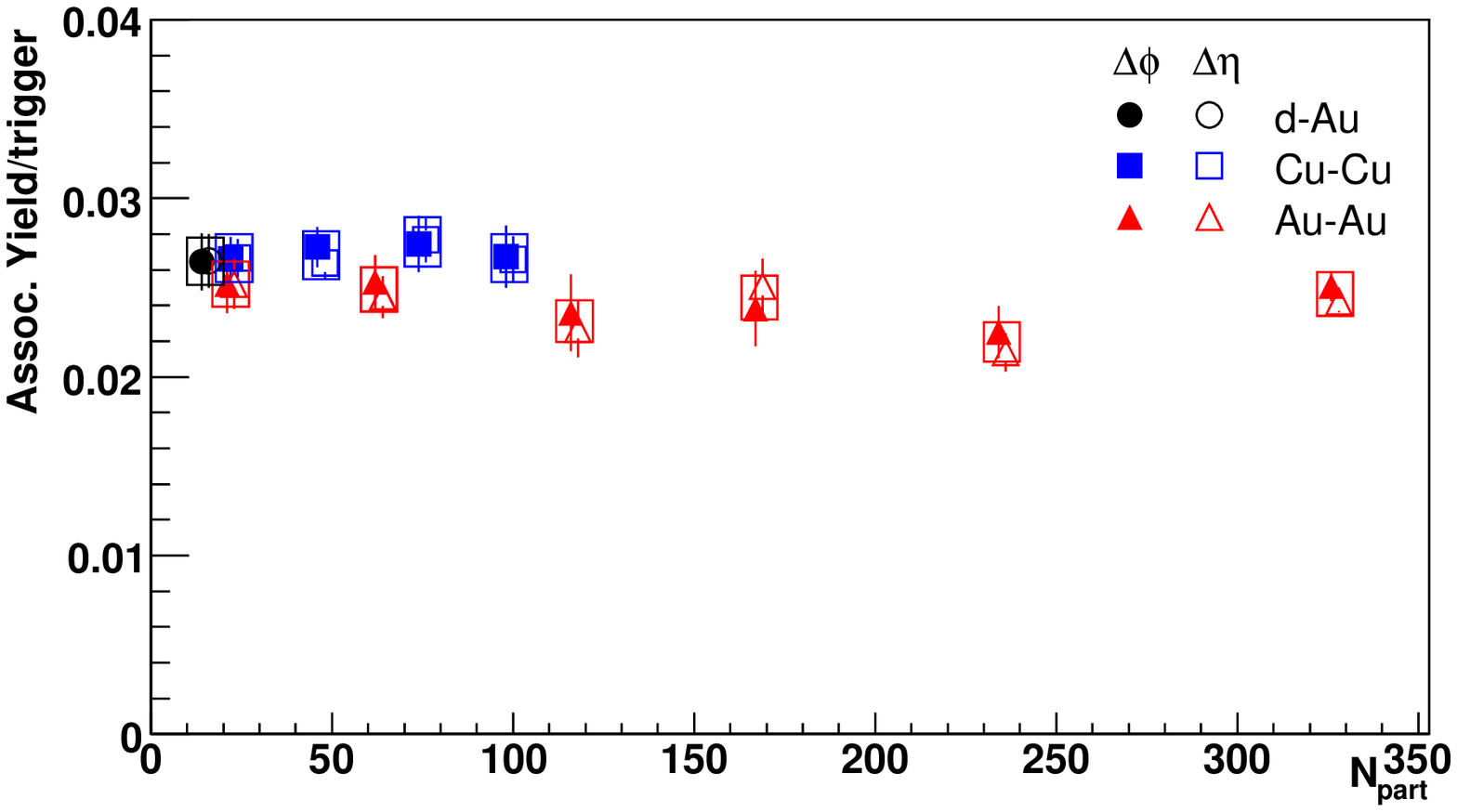}
} \subfloat[]{
\includegraphics[height=55mm,width=0.5\linewidth]{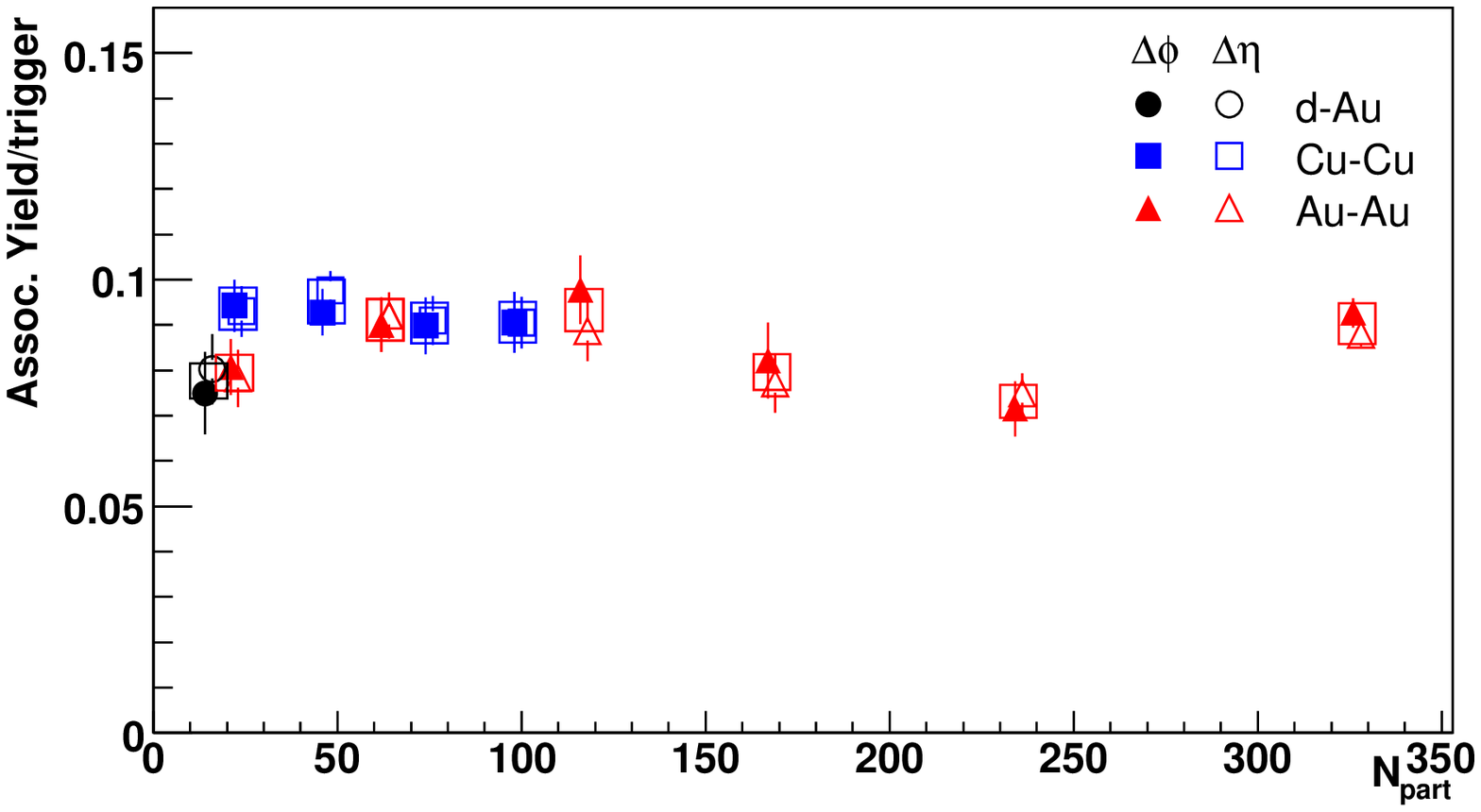}
} \caption{$N_{part}$ dependence of the the near-side associated-particle yield for two
trigger $p_{T}$ ranges: (a) 4~GeV$/c$~$<p_{T}^{trig}<$~6~GeV/$c$, (b) 6~GeV/$c$~$<p_{T}^{trig}<$~10~GeV/$c$. 
For both panels 3~GeV/$c<p_{T}^{assoc}<p_{T}^{trig}$. The hollow symbols are horizontally offset for clarity. }
\label{nearresults}
\end{figure*}

The near-side associated-particle yield, defined as
\begin{equation}
Y^{near}_{AA}\!=\!\!\int^{0.7}_{-0.7}\!\!d(\Delta\eta)\!\!\!\int^{0.78}_{-0.78}\!\!d(\Delta\phi)\frac{1}{N_{trig}}\frac{d^{2}N_{corrected}}{d(\Delta\eta) d(\Delta\phi)}\;
\end{equation}
is presented as a function of number of participant nucleons
($N_{part}$) in Fig.~\ref{nearresults}. The two methods produce
results that are consistent with each other. The Cu+Cu and Au+Au
near-side associated yields are consistent within errors for similar
$N_{part}$. The near-side yields in heavy-ion collisions show no
centrality dependence and within errors agree with those in d+Au, as
seen also in previous studies~\cite{Dan}. The observed independence of
the near-side associated yields on centrality indicates that in this
\pt-range fragmentation is largely unmodified by the presence of the
medium. Note that this does not necessarily imply that those partons
do not lose energy, but rather that they fragment outside the medium
after energy loss. In that case, the energy loss would reduce the
number of trigger hadrons at a given \pt, but not change the
associated particle distribution at intermediate to high \pt. The
enhancement of associated particles at {\it low} \pt{} that has been
reported earlier \cite{ridge1} could then be due to fragments of
radiated gluons.

\begin{figure}
 \centering
  \includegraphics[width=0.8\linewidth,height=5cm]{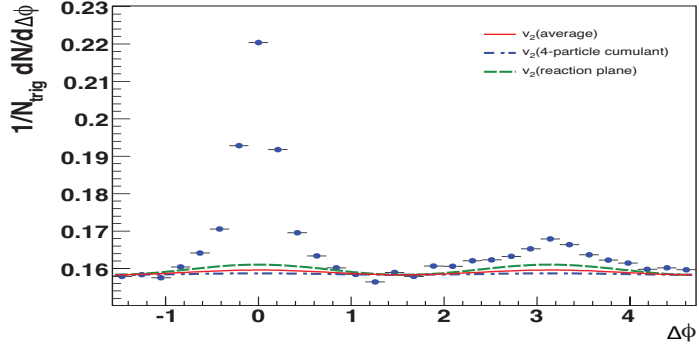}
\caption{$\Delta\phi$ distribution in (0-12\%) central Au+Au collisions used to
extract the away-side yield, 4~GeV/$c$~$<p_{T}^{trig}<$~6~GeV/$c$,
3~GeV/$c<p_{T}^{assoc}<p_{T}^{trig}$, $|\Delta\eta|<1.7$. The triangular pair acceptance correction in $\Delta\eta$ is not applied. The green line represents the elliptic flow modulated background using the values of $v_{2}$ calculated using the reaction plane method, the blue line uses the $v_{2}$ obtained using the 4-particle cumulant method. The red line uses the average value of $v_{2}$.} 
\label{awaycorr}
\end{figure}

The choice of high-$p_{T}$ trigger particles leads to a surface bias in the distribution of hard scattering points~\cite{RenkC75}. The away-side partons have longer path lengths through the medium and therefore will suffer higher energy losses that lead to away-side yield suppression. The study of the away-side yield suppression provides an important tool for determining the energy loss dependence on path length. The away-side associated-particle yield is measured by integrating the associated
hadrons in the region $|\Delta\phi-\pi|<1.3$, covering the azimuthal range of the away-side jet. A background subtraction
is applied to remove the background which is correlated with the
trigger particles through elliptic flow $v_{2}$. The elliptic flow modulated background is described by $dN/d(\Delta\phi)=B\left(1+2\langle
v_{2}^{trig} v_{2}^{assoc}\rangle \cos(2\Delta\phi)\right)$, and is
illustrated in Fig.~\ref{awaycorr} for central collisions. The background is subtracted
using the assumption that there is no jet contribution at the minimum
of the distribution~\cite{phenixZYAM} --- in this case at $|\Delta\phi|\sim$~1.  

\begin{figure*}
\subfloat[]{
\includegraphics[height=55mm,width=0.5\linewidth]{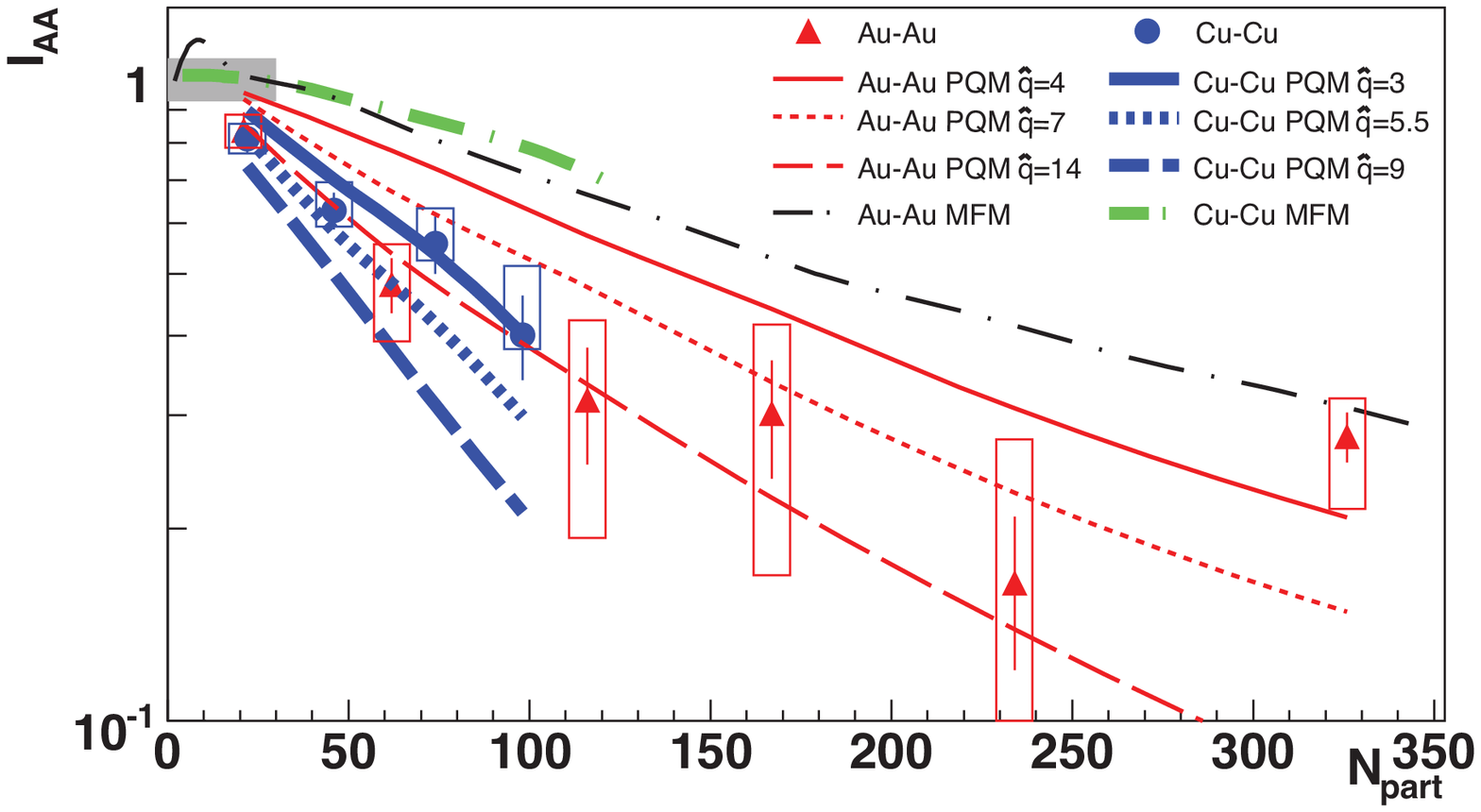}
} \subfloat[]{\includegraphics[height=55mm,width=0.5\linewidth]{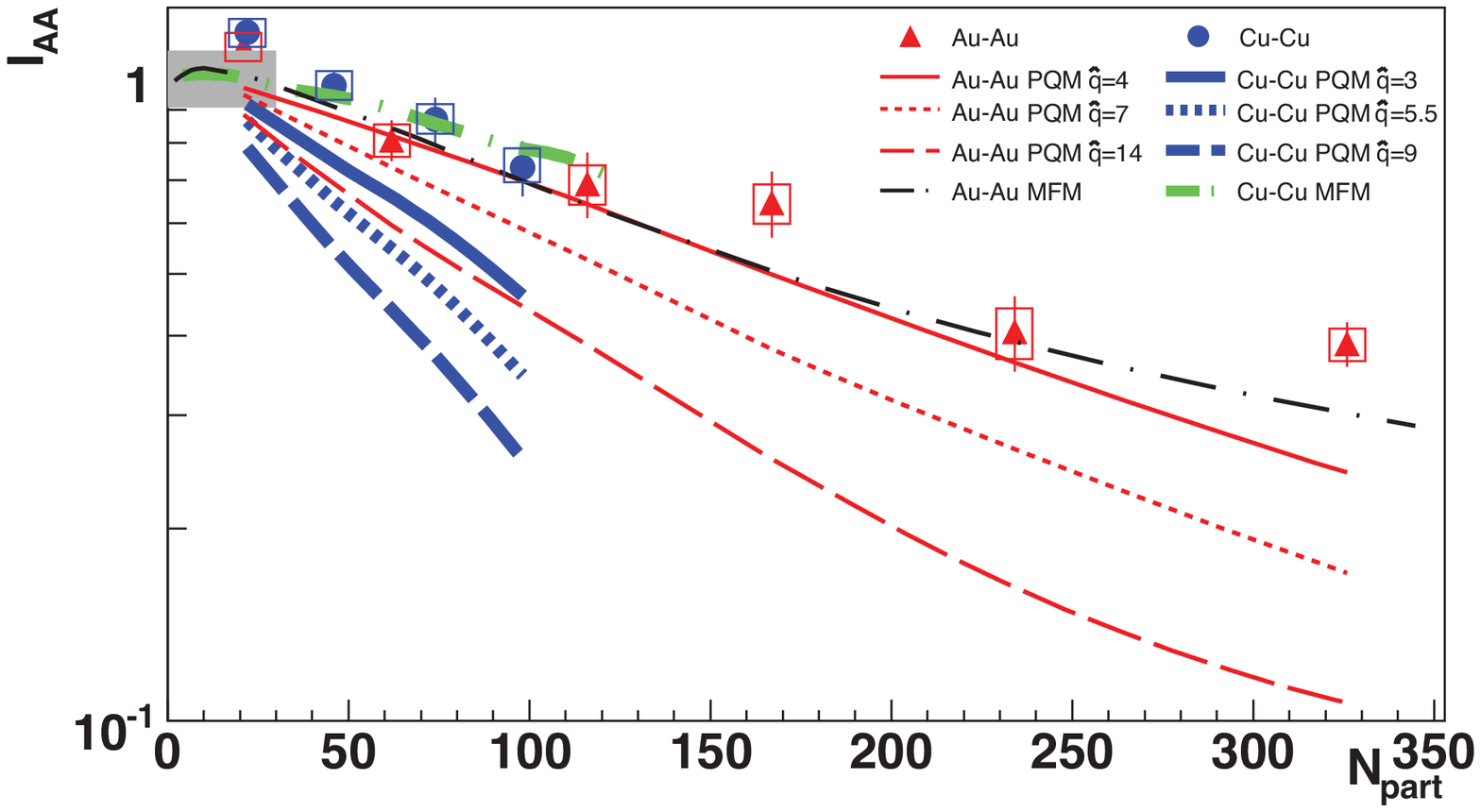}
} \caption{$N_{part}$ dependence of the away-side associated-particle yield for two
trigger $p_{T}$ ranges: (a) 4~GeV/$c$~$<p_{T}^{trig}<$~6~GeV/$c$
(b) 6~GeV/$c$~$<p_{T}^{trig}<$~10~GeV/$c$. For both panels 3~GeV/$c<p_{T}^{assoc}<p_{T}^{trig}$. The error bars represent statistical errors and the boxes represent the point-to-point systematic errors. The gray band represents the correlated error due to the statistical error in the d+Au data. The lines represent calculations in PQM and MFM models. The values of $\hat{q}$ are expressed in GeV$^{2}$/fm.}
\label{awayresults}
\end{figure*}

The amplitude of the background modulation is
given by $\langle v_{2}^{trig} v_{2}^{assoc}\rangle \approx \langle
v_{2}^{trig}\rangle \langle v_{2}^{assoc}\rangle$ which is measured in
STAR using a number of different methods~\cite{flowpaper}. 
For the Au+Au collisions, the nominal value of $v_{2}$ for the
background subtraction was the average between the four-particle
cumulant and the reaction plane measurements of $v_2$. In the Cu+Cu
case, the nominal value is the average between the $v_2$ results obtained using two methods. The first method is the reaction plane method using tracks in the Forward Time Projection Chamber~\cite{ftpc}. The second method uses tracks in the TPC but subtracts the azimuthal correlations in p+p collisions to remove non-flow correlations. The systematic uncertainty associated with the background removal is
estimated in both cases as the difference between the results given by each method
and the nominal value.

The background subtracted away-side yields are used to compute the
suppression factor $I_{AA}=Y_{AA}^{away}/Y_{dAu}^{away}$, where
$Y_{AA(dAu)}^{away}$ is the away-side di-hadron correlation strength in
heavy-ion and d+Au collisions, respectively. Figure~\ref{awayresults} shows the results for $I_{AA}$ as a function of
number of participants for Cu+Cu and Au+Au collisions. The away-side yield suppression increases with
$N_{part}$, as expected. The Cu+Cu results show
a similar suppression ($I_{AA}$) at the same number of participants as the Au+Au
results, despite possible differences in density and path length
distributions.

Figure~\ref{awayresults} also shows two model calculations implementing the same kinematic cuts as our analysis.\footnote{The model calculations use p+p as the reference, which is expected to be equivalent to the d+Au measurement used in the data.} One calculation, the Parton Quenching
Model (PQM)~\cite{Loizides2,Loizides1}, uses the Salgado-Wiedemann
quenching weights~\cite{Salgado} with a Glauber-overlap geometry in
which the local density scales with the local density of binary collisions
$\rho_{coll}$. The other model uses a next-to-leading order QCD
calculation with modified fragmentation functions from a higher-twist
formalism~\cite{XNian1} and a hard-sphere geometry where the density
scales with the local participant density $\rho_{part}$~\cite{XNian2}. We refer to this model as the Modified
Fragmentation Model (MFM). The MFM authors used previous data on the suppression of high-\pt{} away-side yields in central Au+Au collisions~\cite{Dan} to tune their model. The PQM authors present 3 calculations, based on 3 values of $\hat{q}$ in central collisions, indicated by different line styles in the figure.

For the lower trigger selection, 4~GeV/$c$~$ < p_{T}^{trig} <$~6~GeV/$c$, the Modified Fragmentation Model predicts a smaller suppression
than observed in the data, whereas PQM cannot explain Cu+Cu or Au+Au results in a consistent fashion. The disagreement between the models and the data suggests that
the effect of kinematic limits (energy loss cannot be larger than the jet
energy) and non-perturbative effects, which are not explicitly treated in the
model, are significant in this \pt-range. For the higher trigger
\pt{} range, 6~GeV/$c$~$<p_{T}^{trig}<$~10~GeV/$c$, a better agreement between
the data and MFM is observed. There is an obvious difference between the system size dependence in the two models. While MFM obtains  $I_{AA}$ values that are independent on the system at a certain $N_{part}$, PQM shows a clear difference between the two systems for similar $N_{part}$, when using a common scaling of the medium density (represented by line styles in the figure). Further model studies
are needed to clarify whether the different scaling behavior in MFM and PQM
is mainly a result of the different quenching formalisms or rather due
to differences between the medium density models.

\begin{figure}
\centering
  \includegraphics[width=0.8\linewidth,height=9cm]{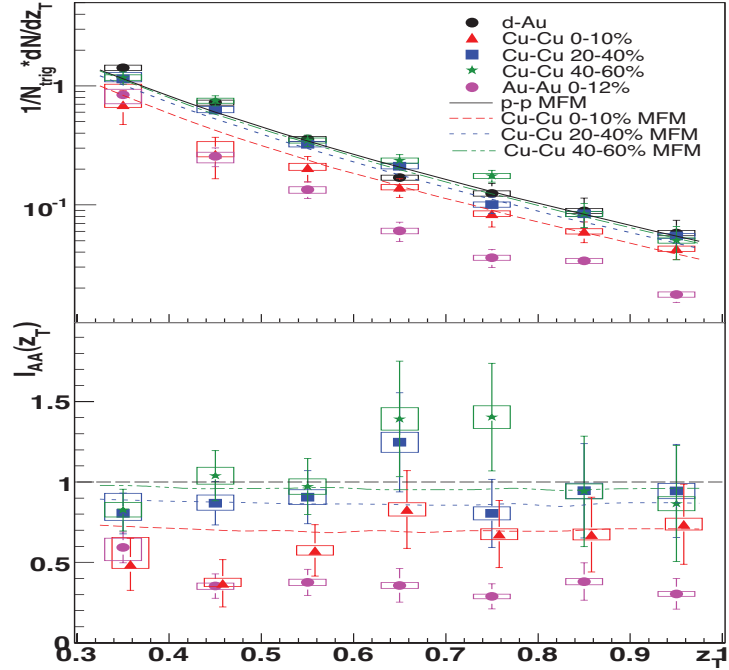}
\caption{Away-side associated particle distribution and $I_{AA}$ for 6~$<p_{T}^{trig}<$10~GeV/$c$. The error bars represent statistical errors and the boxes represent the total systematic errors. The lines represent calculations in MFM model.}
\label{frag}
\end{figure}

In Figs.~\ref{nearresults} and~\ref{awayresults} we have presented
results for a single selection of associated hadrons, \ptassoc~$>$~3~GeV/$c$. A more differential measurement is presented in Fig.~\ref{frag}, which shows the away-side associated yield as a
function of  $z_{T}=p_{T}^{assoc} /
p_{T}^{trig}$. The lower panel of Fig.~\ref{frag} shows the
\zt-dependence of $I_{AA}$. The away-side suppression
is approximately independent of $z_{T}$ in the measured range,
indicating that the momentum distribution of fragments along the jet
axis is not modified by energy loss. A possible explanation of the $z_{T}$-independent $I_{AA}$ is that energy loss is large enough that partons which lose energy have such a soft fragment distribution that they do not contribute significantly to the away-side yield. The remaining away-side yield would then be dominantly from the fraction of partons that lost little or no energy due to a short path length (surface bias, tangential jets) or energy loss fluctuations.  Also shown in Fig.~\ref{frag} are calculations in the Modified Fragmentation Model~\cite{XNian2}, which agree with the results within the present statistical uncertainties. 


In summary, we have presented a systematic study of di-hadron correlations of particles associated with high
transverse momentum trigger hadrons. We have studied the jet-like
correlations on the near-side ($\Delta\phi\sim$~0) and away-side
($\Delta\phi\sim\pi$) for d+Au, Cu+Cu and Au+Au
collisions at $\sqrt{s_{NN}}=200$ GeV/c. Near-side associated yields are equal within the experimental uncertainty for all the
systems studied and independent of the number of participant nucleons
($N_{part}$). Away-side associated yields are suppressed in heavy-ion
collisions with respect to the d+Au reference. The suppression
increases with increasing $N_{part}$ and shows no significant dependence on the
collision system for a given $N_{part}$. The Parton Quenching Model~\cite{Loizides2,Loizides1} 
does not describe the similarity of the away-side yields in the two collision systems at a given $N_{part}$,
while the Modified Fragmentation Model~\cite{XNian2,XNian1} describes
this relatively well for the higher \pt{} triggers. Further comparison of these measurements
to models may allow the extraction of the path length dependence of energy
loss and whether elastic or radiative energy loss
is dominant~\cite{Renkel}. 






We thank the RHIC Operations Group and RCF at BNL, and the NERSC Center 
at LBNL and the resources provided by the Open Science Grid consortium 
for their support. This work was supported in part by the Offices of NP 
and HEP within the U.S. DOE Office of Science, the U.S. NSF, the Sloan 
Foundation, the DFG cluster of excellence `Origin and Structure of the Universe', 
CNRS/IN2P3, RA, RPL, and EMN of France, STFC and EPSRC of the United Kingdom, FAPESP 
of Brazil, the Russian Ministry of Sci. and Tech., the NNSFC, CAS, MoST, 
and MoE of China, IRP and GA of the Czech Republic, FOM of the 
Netherlands, DAE, DST, and CSIR of the Government of India,
the Polish State Committee for Scientific Research,  and the Korea Sci. \& Eng. Foundation.

\end{document}